\documentclass[runningheads]{llncs}
\usepackage[T1]{fontenc}
\usepackage{booktabs}
\usepackage{graphicx}
\usepackage{array}
\newcolumntype{C}[1]{>{\centering\arraybackslash}p{#1}}
\usepackage{amsmath}
\usepackage{amssymb}
\usepackage{url}
\begin{document}
\title{Learning AI Without a STEM Background: Mixed-Methods Evidence from a Diverse, Mixed-Cohort AIED Program}
\titlerunning{Learning AI without a STEM Background}
\author{Valentina Kuskova\orcidID{0000-0003-4716-2544} \and
Dmitry Zaytsev\orcidID{0000-0002-0902-3896} \and
Richard Patrick Johnson\orcidID{0000-0002-1550-6325}}
\authorrunning{V. Kuskova et al.}
\institute{Lucy Family Institute for Data \& Society \\
University of Notre Dame, Notre Dame, IN, USA}
\maketitle
\begin{abstract}
Despite growing interest in AI education, most AIED initiatives remain narrowly targeted toward STEM-prepared students, limiting participation by non-STEM learners and adults seeking to engage with AI in public-interest, policy, or workforce contexts. This paper presents and evaluates an NSF-funded, innovative mixed-cohort AI education model that intentionally integrates non-STEM undergraduates and adult learners into a shared learning environment centered on ethical reasoning, socio-technical judgment, and applied AI literacy rather than technical proficiency alone.
Drawing on mixed-methods data from course surveys, open-ended reflections, and educator reports, we examine learners' academic agency, confidence navigating AI concepts, critical engagement with ethical tradeoffs, and perceived expansion of postsecondary and career trajectories. Quantitative results indicate significant gains in confidence and perceived relevance of AI across cohorts' participants, while qualitative analyses reveal a consistent emphasis on responsibility, judgment, and contextual reasoning over technical mastery. Instructors and near-peer mentors corroborated high levels of engagement and productive challenge, particularly in dialogic and scenario-based learning activities.
Our findings suggest that human-centered instructional supports, such as ethical scaffolding, mentorship, and structured discussion, are essential components of equitable AI education, especially in heterogeneous and non-traditional learner populations. We argue that ethical judgment should be treated as a core learning outcome in AIED alongside AI literacy, and we offer design implications for expanding access to AI education in policy-relevant and workforce-adjacent contexts.

\keywords{Innovative Mixed-Cohort AI Education  \and Non-STEM AIED \and Workforce-Adjacent AI Learning.}
\end{abstract}

\section{Introduction: The Access Problem in AI Education}
Artificial intelligence is increasingly embedded across work, governance, and everyday life, reshaping how decisions are made, resources are allocated, and opportunities are distributed \cite{Dunleavy}. As a result, AI literacy is now widely recognized not merely as a technical skill, but as a form of civic, professional, and ethical competence \cite{Long}. Yet access to AI education remains uneven \cite{Ahmed}. Most formal AI learning opportunities are designed for STEM-prepared students \cite{Nguyen}, rely heavily on coding proficiency \cite{Amanuel}, or target early-career learners \cite{van der Linde}, systematically excluding non-STEM students and adults seeking to engage with AI in policy, public-interest, or workforce contexts.

These access barriers have consequences beyond pedagogy. Restricting AI education to technically trained populations reinforces existing inequities \cite{Li} in technological participation and decision-making. Individuals without accessible AI education are less equipped to critically evaluate automated systems \cite{Faccia}, participate in AI-mediated organizational processes \cite{Florea}, or adapt to labor markets increasingly shaped by data-driven tools \cite{He}. For adult learners embedded in professional roles \cite{Kim}, the absence of rigorous, accessible AI education pathways limits opportunities for upskilling, mobility, and ethical leadership.

Prevailing AI education models often intensify these challenges. Coding-centric curricula \cite{Lim} could prioritize implementation over judgment, degree-bound programs impose time and financial constraints \cite{Osetskyi}, and youth-focused initiatives may overlook the learning needs of experienced professionals \cite{Poquet}. While valuable in their own right, these approaches leave unresolved how AI education might expand access without sacrificing rigor, ethical depth, or real-world relevance.

This paper addresses that gap by examining a mixed-cohort AI education model that brings together non-STEM undergraduates and adult learners in a shared learning environment. The study is guided by two research questions: (1) How can AI education be designed to expand access while maintaining rigor? and (2) What learning outcomes emerge in mixed-cohort, non-traditional AI classrooms? The program under study---referred to here as \textit{Data Crossings}---was supported by the National Science Foundation (NSF).\footnote{The full NSF project title is withheld for peer review to avoid author identification.} Drawing on empirical evidence from this year-long program, we examine how learners develop academic agency, confidence engaging with AI concepts, and the capacity to reason about ethical and socio-technical tradeoffs.

The paper makes three contributions. First, it introduces a mixed-cohort AIED design that integrates learners across disciplines, ages, and career stages. Second, it presents mixed-methods evidence from a sustained, evaluated program rather than a short-term intervention. Third, it reframes ethical judgment and academic agency, rather than technical mastery alone, as core outcomes of AI education, with implications for AIED research and policy-oriented program design.

\section{Related Work}
\textbf{AI Education and Participation Gaps.} Research in AI education and computing education has consistently documented disparities in participation across gender \cite{Conde}, socioeconomic status \cite{Ahmed}, disciplinary background \cite{Vindigni}, and educational stage \cite{Pedro}. While many initiatives aim to broaden participation, they often focus on pre-college pipelines \cite{Shailja} or undergraduate STEM majors \cite{Hasan}, leaving non-STEM students and adult learners underrepresented. These gaps are particularly pronounced in advanced or applied AI learning contexts \cite{Cai}, where prerequisites and assumptions about prior knowledge function as de facto exclusion mechanisms. As AI systems increasingly influence public policy, healthcare, education, and civic life, these participation gaps raise concerns about who is equipped to understand, question, and govern AI technologies.

\textbf{AI Literacy vs.\ Technical Proficiency.} A growing body of literature distinguishes between AI literacy and technical proficiency, arguing that meaningful engagement with AI does not require deep expertise in programming or model development \cite{Mallik}. AI literacy frameworks emphasize conceptual understanding \cite{Yu}, critical interpretation of AI outputs \cite{Chiu}, awareness of limitations and biases \cite{Oyetade}, and the ability to situate AI systems within social and organizational contexts \cite{Makarius}. However, many AI education programs continue to conflate literacy with technical skill acquisition, privileging coding fluency over interpretive and ethical capacities. This conflation disproportionately disadvantages learners whose goals involve policy analysis, management, communication, or domain-specific application rather than technical implementation.

\textbf{Adult Learning and Workforce-Adjacent AI Education.} Adult learning research highlights that adults bring prior knowledge, professional identity, and goal-oriented motivations that differ substantially from traditional students \cite{Kasworm}. Workforce-adjacent AI education programs have emerged in response to rapid technological change, yet many rely on self-paced online formats or narrowly defined upskilling modules \cite{Oschinski}. These approaches often struggle with retention, depth of learning, and ethical engagement. Moreover, adult learners are rarely studied as participants in AI education research, particularly in environments that position them as co-learners rather than remedial or peripheral participants.

\textbf{Ethics, Responsibility, and Judgment in AIED.} Ethical considerations are increasingly recognized as central to AI education, with scholars advocating for the integration of fairness, accountability, transparency, and social impact into AI curricula. However, ethical AI education is frequently treated as an add-on \cite{Garrett}: isolated modules or discussions appended to otherwise technical courses. Less attention has been paid to how ethical judgment develops through sustained practice, dialogue, and exposure to real-world tradeoffs, particularly among learners without technical backgrounds. Understanding ethics as a form of reasoning and judgment, rather than rule-following, remains an underexplored dimension of AIED.

\textbf{Mixed-Cohort and Peer-Learning Models: What's Missing.} Peer learning and cohort-based models are well established in STEM education and professional training \cite{Pradyutha}, with evidence that collaboration and social belonging support retention and learning. Existing mixed-cohort studies, however, primarily focus on disciplinary diversity among students at similar educational stages \cite{Weerakkody}. Mixed cohorts that span both age and career stage, integrating adult professionals with undergraduate learners, remain largely absent from AIED research. This absence limits our understanding of how cross-generational and cross-experiential learning might shape AI education outcomes. Addressing this gap is essential for designing inclusive AI education models that reflect the diversity of real-world AI stakeholders.

\section{Program Context and Instructional Design: The Data Crossings Model}

The Data Crossings program was developed in response to NSF priorities articulated in the Experiential Learning for Emerging and Novel Technologies (ExLENT) initiative, which emphasize inclusive STEM pathways, experiential learning, and workforce development aligned with societal needs \cite{ExLENT}. Conceived as a year-long exploratory program, Data Crossings seeks to broaden access to AI and data science education for learners typically excluded from traditional STEM pipelines, while maintaining intellectual rigor, ethical grounding, and applied relevance.

A defining feature of the program is its mixed-cohort structure, which intentionally integrates non-STEM undergraduate students and adult learners with prior professional or domain expertise. Undergraduate participants are drawn from the humanities, social sciences, arts, and business disciplines and seek engagement with AI without committing to formal STEM degree pathways. Adult learners include working professionals and career-transitioning individuals interested in applying AI in organizational, policy, or community contexts. These populations are deliberately placed in shared instructional and project-based environments to support peer learning across age, experience, and disciplinary boundaries.

\subsection{Program Structure}

Data Crossings is organized around four integrated components delivered over an academic year. First, an Explorations course provides a no-code introduction to AI and data science, including data literacy, model interpretation, design thinking, and ethical considerations. Second, Experiential Learning Studios (ELS) run in parallel with the course, offering structured, collaborative spaces for applying concepts to real-world problems. Third, participants complete team-based capstone projects grounded in societal or industry-relevant challenges. Fourth, the program culminates in in-situ experiential learning, including internships or participation in the Research and Industry Scholarly Exchange Fellowship Program (RAISEuP), enabling learners to apply AI concepts in authentic organizational settings. In addition, the program includes a series of workshops focused on developing transferable ``superskills'' \cite{superskills}, such as communication, collaboration, and professional judgment.

\subsection{Instructional Design and Mixed-Cohort Learning}

Instruction emphasizes dialogic and experiential pedagogies rather than lecture-based delivery. Biweekly sessions combine guided discussion with hands-on activities using no-code and low-code tools, allowing participants to explore real datasets, interpret model outputs, and examine ethical tradeoffs without focusing on programming syntax or algorithmic implementation. Scenario- and case-based learning plays a central role, particularly in ethical instruction, positioning ethical judgment as an active, practice-based capacity.

The year-long structure enables progressive scaffolding of analytic complexity and ethical responsibility, culminating in capstone projects that integrate technical, contextual, and ethical dimensions. Mixed-cohort learning is embedded throughout the curriculum: undergraduate students and adult learners collaborate in shared instructional spaces and project teams, leveraging complementary strengths. Each project group includes at least one adult learner alongside two to three undergraduate students, with peer mentorship structured as reciprocal rather than hierarchical, reflecting a commitment to epistemic equity and recognition of multiple forms of expertise within AI education.

\textbf{Evaluation.} Program evaluation is conducted by the Synergy Evaluation Institute at the University of Tennessee, Knoxville, using a mixed-methods design aligned with NSF priorities. The evaluation examines implementation fidelity, quality of the mixed-cohort learning experience, development of AI literacy and ethical reasoning, and sustainability of cross-sector partnerships. Data sources include participant surveys, interviews, capstone artifacts and rubrics, project documentation, and stakeholder feedback, enabling triangulation across learner, instructor, and partner perspectives.

\section{Data and Methods}

This study employs a mixed-methods evaluation design to examine how participants experienced and learned within the Data Crossings mixed-cohort AI education program. The analytic approach integrates quantitative and qualitative data sources to capture changes in learner confidence, agency, and ethical reasoning, as well as to triangulate participant self-reports with educator perspectives. The methods are intentionally aligned with the program's human-centered and equity-oriented design, emphasizing interpretive validity over narrowly defined technical performance metrics.

\subsection{Participants}

Participants were drawn from the first implementation cohort of the Data Crossings program, which intentionally integrated non-STEM undergraduates and adult learners with prior professional or domain expertise. More than 200 individuals applied in the first call; 19 were selected (9.45\% acceptance rate), including 12 undergraduate students and 7 adult learners. This mixed-cohort composition was a deliberate design feature rather than a sampling artifact, yielding substantial variation in age, educational background, and prior exposure to AI and data concepts.

Undergraduate applicants primarily represented the humanities and arts (approximately $20\%$), social sciences and policy ($\approx$35\%), and non-STEM business and economics ($\approx$30\%), with smaller representation from architecture and design ($\approx$8\%). Only a small share of applicants came from STEM fields such as computer science ($\approx$7\%). Adult learners included working professionals and career-transitioning individuals seeking to apply AI and data science in domain-specific contexts.

Selection emphasized programmatic fit over technical proficiency. Review criteria prioritized motivation to engage with AI in non-technical domains, interest in ethical and socially situated applications, readiness for mixed-cohort learning, and alignment with the program's exploratory, experiential design. In accordance with NSF requirements, applicants with formal STEM training in computing disciplines were excluded, ensuring that the cohort reflected populations typically underserved by traditional AI and data science programs.

\subsection{Data Sources and Analytic Approach}

Multiple data sources were collected over the course of the program to support both formative and summative analysis. Participant surveys were administered at multiple points in the program, including immediately upon entering the program and multiple times throughout the program. Survey items assessed self-reported confidence engaging with AI concepts, perceived relevance of AI to participants' academic or professional goals, academic agency, and comfort reasoning about ethical implications of AI use. Likert-style items were supplemented with open-ended prompts to allow participants to elaborate on their experiences.

Open-ended reflections were collected through written responses embedded within course activities and end-of-course surveys. These reflections prompted participants to describe how their understanding of AI evolved, how they approached ethical tradeoffs, and how the mixed-cohort structure shaped their learning.

Instructor and teaching fellow surveys provided an additional perspective on participant engagement, challenge level, and learning dynamics. These instruments focused on observed growth in confidence, quality of discussion, and participants' ability to articulate ethical and contextual considerations when working with AI tools.

\textbf{Analytic Approach.} The analysis followed a convergent mixed-methods strategy, in which quantitative and qualitative data were analyzed separately and then integrated to support interpretation.

Quantitative survey data were analyzed descriptively to examine patterns of change over time in participant confidence, perceived relevance of AI, and academic agency. Given the exploratory nature of the program and the intentionally small cohort size, analyses emphasize directional trends and practical significance rather than inferential generalization. Comparisons across learner subgroups (e.g., undergraduate versus adult learner) were examined where appropriate to identify convergent or divergent patterns.

\subsection{Text Analysis Methods}

Applicant narratives were analyzed using a hybrid computational--qualitative NLP approach appropriate for medium-sized educational text corpora. The method combines unsupervised topic modeling with human-in-the-loop coding to balance reproducibility and interpretive validity \cite{DIgnazio,Nelson}. Texts were preprocessed using standard procedures (lowercasing, tokenization, lemmatization, stopword removal), with sentence segmentation retained to capture multiple motivations within individual responses.

Initial theme discovery used unsupervised topic modeling, conceptually aligned with Latent Dirichlet Allocation and embedding-based clustering methods such as BERTopic, to identify latent semantic structure through word co-occurrence and contextual similarity \cite{Blei,Grootendorst}. Algorithmic outputs were treated as exploratory and consolidated through qualitative coding into theoretically meaningful themes grounded in AI education and equity literatures. Coding was conducted at the idea level, allowing multi-label assignment consistent with best practices for educational narrative analysis \cite{Braun,Nelson}. Theme frequencies reflect the proportion of applicants mentioning a theme at least once; percentages therefore do not sum to $100\%$. This approach supports analytic rigor while remaining sensitive to contextual and ethical nuance \cite{DIgnazio,Zhai}.

\section{Results: Early Findings from Program Entry and Initial Surveys}
Because the Data Crossings program is ongoing at the time of writing, comprehensive post-program learning outcomes for the first cohort are not yet available. Accordingly, this section reports early findings drawn from application materials and surveys administered during the initial phase of the program. These data provide insight into who the program is reaching, participants' motivations and expectations, and baseline orientations toward AI, ethics, and learning. Rather than evaluating outcomes, the analyses reported here establish the conditions under which learning is expected to occur and assess alignment between program design goals and participant needs.

\subsection{Motivations for Participation and Access Pathways}
Analysis of application narratives ($n=201$; Table~\ref{tab:applicant_nlp_themes}) reveals strong convergence in motivations despite wide disciplinary diversity. The dominant theme is ethical and responsible use of AI, with many applicants expressing concerns about bias, misuse, and social consequences. Notably, these ethical considerations often precede expressions of technical confidence, suggesting that ethical awareness is an entry condition for non-traditional AI learners.

\begin{table}[htbp]
\centering
\caption{Thematic Analysis of Applicant Motivations for Participating in the ExLENT Data Crossings Program (Sample)}
\label{tab:applicant_nlp_themes}
\begin{tabular}{|p{3cm}|c|p{7.2cm}|}
\hline
\textbf{Theme} & \textbf{Frequency} & \textbf{Illustrative Description} \\
\hline
Ethical and Responsible Use of AI & 68\% &
Desire to understand, govern, and apply AI responsibly; concerns about bias, misuse, accountability, and social harm. \\
\hline
AI and Data for Social Good & 65\% &
Motivation to apply AI and data science to public policy, nonprofits, healthcare, education, sustainability, and community impact. \\
\hline
Non-STEM Access and Inclusion & 61\% &
Explicit identification as non-STEM learners; emphasis on accessibility, no-coding pathways, and overcoming structural barriers to AI education. \\
\hline
Career Preparation and Workforce Readiness & 59\% &
Interest in AI literacy as essential for future careers across law, business, policy, finance, consulting, and public service. \\
\hline
Interdisciplinary Learning and Translation & 56\% &
Desire to bridge humanities, social sciences, and professional domains with AI and data tools; focus on translation rather than model building. \\
\hline
\end{tabular}
\end{table}

Applicants also consistently frame AI and data science as tools for social good, spanning policy, nonprofit work, healthcare, education, and sustainability. Technical literacy is rarely described as an end in itself; instead, applicants emphasize interpretation, judgment, and decision-making in real-world contexts.

A third recurring pattern concerns access barriers for non-STEM learners. Many applicants describe traditional AI pathways as inaccessible due to coding prerequisites or curricular constraints, while identifying the program's no-coding design and structured support as enabling participation. Relatedly, applicants emphasize workforce relevance across diverse fields, positioning themselves as users and evaluators of AI systems rather than developers.

Motivations are broadly similar across undergraduate and adult applicants. Undergraduates tend to emphasize exploration and confidence-building, while adult learners emphasize applicability to professional contexts. Across both groups, applicants express high motivation paired with limited prior confidence, underscoring the importance of accessible, experiential, and human-centered AI education.

\subsection{Baseline Confidence, Agency, and Ethical Orientation}

Early survey data indicate that participants entered the program with low to moderate confidence in their ability to engage with AI concepts, despite high levels of interest. Many respondents reported discomfort interpreting AI outputs or assessing system limitations, reinforcing the notion that confidence, rather than motivation, represents a primary barrier to participation in AI education.

Within one month of program participation, the level of confidence has noticeably changed. Table~\ref{tab:ai_comfort_transition} presents the transition table for the responses to the question ``How confident are you using AI technology?'' To further quantify changes in self-reported comfort with AI technology, we computed the expected value shift implied by the transition matrix as follows:

\[
\mathbb{E}[\Delta]
= \frac{1}{5} \sum_{i=1}^{5}
\left( \mathbb{E}\!\left[ C_{\text{later}} \mid C_{\text{initial}} = i \right] - i \right)
\]

Analysis of the transition matrix indicates a clear increase in self-reported comfort with AI technology. Participants who began with low or moderate comfort showed the largest upward movement, while those entering with higher comfort levels largely remained stable. Minor downward shifts among the most confident respondents appear consistent with calibration rather than disengagement. Importantly, no participants transitioned to extreme discomfort, and downward movement was rare.

Expected value analysis reinforces this pattern. Conditional on starting level, respondents exhibited positive gains at all but the highest comfort category, with the largest increases among those initially least comfortable. Averaged across starting levels, the overall expected shift was positive ($\mathbb{E}[\Delta] = +0.90$), indicating a substantial net increase in comfort with AI technology.

Qualitative responses further suggest that participants valued learning environments that legitimize questioning and uncertainty. This emphasis on academic agency---defined as confidence in engaging, asking questions, and forming judgments---aligns with the program's instructional design and provides a foundation for subsequent analysis of learning trajectories.

\begin{table}[t]
\centering
\caption{Transition Matrix of Self-Reported Comfort Using AI Technology}
\label{tab:ai_comfort_transition}
\setlength{\tabcolsep}{8pt}
\begin{tabular}{C{1.5cm}cccccC{1.5cm}C{1.5cm}}
\hline
\textbf{Initial Comfort} & \multicolumn{5}{c}{\textbf{Comfort Level at 1 Month}} & \textbf{Expected Later Comfort} & \textbf{Expected Value Shift}\\
\cline{2-6}
 & 1 & 2 & 3 & 4 & 5 & & \\
\hline
1 & 0\% & 0\% & 0\% & 50\% & 50\% & 4.50 & +3.50\\
\hline
2 & 0\% & 25\% & 25\% & 50\% & 0\% & 3.25 & +1.25\\
\hline
3 & 0\% & 0\% & 22\% & 67\% & 11\% & 3.89 & +0.89\\
\hline
4 & 0\% & 0\% & 0\% & 100\% & 0\% & 4.00 & 0.00\\
\hline
5 & 0\% & 6\% & 17\% & 67\% & 11\% & 3.86 & -1.14\\
\hline
\end{tabular}
\\[4pt]
{\footnotesize\textit{Note.} Both surveys were conducted on a 1--5 Likert scale, where ``1'' meant ``not at all confident'' and ``5'' meant ``completely confident.''}
\end{table}

\subsection{Early Signals of Design Alignment}

Although outcome data are not yet available, early indicators suggest strong alignment between participant expectations and program design principles. Participants consistently valued dialogic learning, mixed-experience collaboration, and scenario-based engagement in early course interactions. Instructor and teaching fellow observations corroborate these self-reports, noting high levels of participation in discussions and willingness to articulate ethical tradeoffs even in the absence of technical certainty. These early signals support the plausibility of the program's theory of change: that accessible, human-centered AI education can engage diverse learners meaningfully from the outset, creating conditions conducive to deeper learning and ethical reasoning as the program progresses.

\section{Discussion and Implications for AIED}

The early findings reported here illuminate not finalized learning outcomes, but the conditions under which equitable AI education becomes possible. Interpreted through this lens, the Data Crossings program provides evidence that mixed-cohort, ethics-centered AI education can expand access without diluting rigor, while fostering forms of engagement often absent from conventional AI learning environments.

\textbf{Mixed-Cohort Learning as a Design Asset.} Contrary to concerns that heterogeneous cohorts fragment instruction, the mixed-cohort design functioned as an epistemic asset. Non-STEM undergraduates benefited from adult learners' domain expertise and professional judgment, while adult learners engaged with reflective and analytic practices less common in workplace training. This reciprocal dynamic normalized uncertainty and broadened the range of perspectives applied to AI-related problems, aligning more closely with real-world AI use, where diverse stakeholders must collaboratively interpret and govern systems.

\textbf{Ethical Judgment as a Core AIED Outcome.} Participants entered the program with strong ethical concerns about AI but limited confidence in their ability to reason about these concerns systematically. Early engagement demonstrates that ethical judgment is learnable and observable, even in the absence of technical mastery. Participants articulated tradeoffs, questioned assumptions, and justified decisions through discussion and case-based reasoning. These findings challenge models that postpone ethics until after technical competence and instead support positioning ethical reasoning as a foundational AIED outcome that can scaffold later technical learning.

\textbf{Adult Learners and Epistemic Equity.} Adult participants consistently framed themselves as professionals seeking tools to interpret and responsibly deploy AI, rather than as novices requiring remediation. Their contributions, grounded in lived experience with organizational constraints and real-world consequences, expanded the analytic scope of learning beyond technical optimization. This suggests that adult participation can elevate the intellectual and ethical depth of AI education and underscores the importance of treating adult learners as epistemic contributors within AIED designs.

\textbf{Human Facilitation, Assessment, and Policy Implications.} A central insight from the early data is the role of human facilitation in enabling equitable participation. Guided discussion, structured reflection, and peer interaction created space for questioning and ethical reasoning without penalizing technical uncertainty. In contrast to self-paced online courses, coding-first bootcamps, and credential-gated programs, this model suggests that facilitation is not ancillary but a core equity mechanism in AI education.

These findings have direct implications for AIED design and assessment. Programs seeking to broaden participation should consider mixed-cohort structures supported by intentional scaffolding and facilitation. Assessment frameworks should extend beyond technical performance to capture ethical reasoning, interpretive competence, and academic agency---capacities central to responsible AI use yet rarely measured in prevailing AIED evaluations.

From a policy and workforce perspective, expanding AI literacy requires investment in accessible, facilitated, and non-degree-bound pathways, particularly for adult learners and professionals. Although implemented within a single institutional context, the design principles underlying Data Crossings---mixed cohorts, ethical scaffolding, no coding prerequisites, and experiential learning---are transferable across institutions and sectors, offering pathways to scale inclusive AI education without reproducing traditional STEM exclusivity.

\section{Limitations and Future Work}

This study has several limitations. In line with NSF program setup, the cohort size is modest, and participants self-select (through the application process) into the program, which limits claims of generalizability. The analysis relies on application materials and early survey data, capturing orientations and expectations rather than completed learning outcomes. Additionally, findings are drawn from a single program situated within a specific institutional and regional context.

Future work will address these limitations through longitudinal analysis of post-program survey data, capstone project artifacts, and follow-up interviews with participants and employers. Planned analyses will examine changes in confidence, ethical reasoning, and career trajectories over time, as well as differences in learning experiences across cohort subgroups. Comparative studies across institutions and program designs would further strengthen understanding of mixed-cohort AI education models.

\section{Conclusion}

As AI systems increasingly shape social, economic, and political life, the question of who AI education is for becomes a matter of equity, governance, and public responsibility. This paper argues that expanding AI education requires more than scaling existing technical curricula; it requires rethinking design assumptions about learners, learning goals, and legitimate forms of expertise.

The Data Crossings program demonstrates that mixed-cohort, ethics-centered AI education is not only feasible but promising. By treating ethical judgment and academic agency as core outcomes, positioning adult learners as epistemic contributors, and centering human facilitation as an equity mechanism, the program offers an alternative model for inclusive AI education.

We conclude by calling for broader adoption and systematic study of mixed-cohort AI education within the AIED community. Doing so is essential for preparing a diverse set of learners---not just future developers, but future decision-makers---to engage responsibly with AI technologies that increasingly shape collective life.

\begin{credits}
\subsubsection{\ackname} This material is based upon work supported by the National Science Foundation under Grant Number XXXXXXX.\footnote{To be disclosed after peer review.}

\subsubsection{\discintname}
The authors have no competing interests to declare that are relevant to the content of this article.
\end{credits}

\end{document}